# Wireless Sensor Networks Localization Methods: Multidimensional Scaling vs. Semidefinite Programming Approach

Biljana Stojkoska, Ilinka Ivanoska, Danco Davcev,

[1] Faculty of Electrical Engineering and Information Technologies – Skopje, Karpoš II bb, 1000 Skopje, Macedonia
{biles@feit.ukim.edu.mk, ilinka_iv@yahooo.com, etfdav@feit.ukim.edu.mk}

**Abstract.** With the recent development of technology, wireless sensor networks are becoming an important part of many applications such as health and medical applications, military applications, agriculture monitoring, home and office applications, environmental monitoring, etc. Knowing the location of a sensor is important, but GPS receivers and sophisticated sensors are too expensive and require processing power. Therefore, the localization wireless sensor network problem is a growing field of interest. The aim of this paper is to give a comparison of wireless sensor network localization methods, and therefore, multidimensional scaling and semidefinite programming are chosen for this research. Multidimensional scaling is a simple mathematical technique widely-discussed that solves the wireless sensor networks localization problem. In contrast, semidefinite programming is a relatively new field of optimization with a growing use, although being more complex. In this paper, using extensive simulations, a detailed overview of these two approaches is given, regarding different network topologies, various network parameters and performance issues. The performances of both techniques are highly satisfactory and estimation errors are minimal.

**Keywords:** Wireless Sensor Networks, Semidefinite programming, multi-dimensional scaling, localization techniques

## 1 Introduction

New technologies bring new possibilities, however, in the same time new questions are being opened. The area of wireless sensor networks solves a great amount of new problems. A wireless sensor network (WSN) is a network consisting of distributed sensor devices that cooperatively monitor physical or environmental conditions at different locations. The development of wireless sensor networks was originally motivated by military applications. However, wireless sensor networks are now used in many industrial and civilian application areas, including industrial process monitoring and control, machine health monitoring, environment and habitat monitoring, healthcare applications and traffic control. Today, wireless sensor networks has become a key technology for different types of smart environments, and the aim is to enable the application of wireless sensor networks for a wide range of industrial problems. Wireless networks are of particular importance when a large number of sensor nodes have to be deployed.

A fundamental problem in wireless sensor networks is localization i.e. the determination of the geographical locations of sensors. Localization is a challenge when dealing with wireless sensor nodes, and a problem which has been studied for many years [1]. Nodes can be equipped with a Global Positioning System (GPS), but this is a costly solution in terms of money and power consumption. The localization issue is important where there is an uncertainty about some positioning. If the sensor network is used for monitoring the temperature in a remote forest, nodes may be deployed from an airplane and the precise location of most sensors may be unknown. An effective localization algorithm can then use all the available information from the nodes to compute all the positions.

Most existing localization algorithms were designed to work well in wireless sensor networks. The performance of localization algorithms depend on critical sensor network parameters, such as the radio range, the network topology i.e. the density of nodes, the anchor-to-node ratio, and it is important that the solution gives adequate performance over a range of reasonable parameter values.

In this paper we give an overview of two completely different localization approaches: Multidimensional scaling and Semidefinite programming. We present analysis and simulations of the algorithms, demonstrating the accuracy compared to each other, regarding different sensor network parameters.

The Multidimensional scaling approach is an algorithm using connectivity information for computing the nodes' localization with the help of some linear transformations [2]. The MDS-MAP algorithm first uses connectivity to roughly estimate the distance between each pair of nodes, then, multidimensional scaling (MDS) is used to find possible node locations that fit the estimations, and finally, it is optimized by using the anchors positions [3]. In section 2 we describe the classical MDS approach used in the simulations.

Section 3 describes the Semidefinite programming (SDP) relaxation based method for the position estimation problem in sensor networks [4][5]. The basic idea behind the technique is to convert the nonconvex quadratic distance constraints into convex constraints by introducing a relaxation to remove the quadratic term in the formulation. The solving of the connection convex constraint is by using techniques of linear programming.

In Section 4 we show results on both simulated algorithms with a discussion and a comparison of the two proposed methods and finally, Section 5 concludes the paper.

## 2 Multidimensional Scaling

First we will give a mathematical model of the wireless sensor network localization problem. In all the localization approaches the network is modeled by a graph $G = (V, E)$, where $V$ is the set of nodes some with known positions in the Euclidean space $R^{\dim}$ and $E$ is the set of edges defined by the network topology (connectivity). In one case a set of weights $\{d_{ij} : (i, j) \in E\}$ on the graph's edges is given, representing the (estimated) distances between the corresponding nodes. The problem is then to place all nodes in such a way that the Euclidean distance between every pair of nodes $v$ and $w$, where $(v, w) \in E$, equals $d_{vw}$. In the other case, $d_{ij}$ are not given a special value. It is only assumed that $d_{ij} < R$, where $R$ is the range of the transmitter of a wireless sensor node. To describe the positions of the nodes of the network, we form a corresponding matrix and to store the available distance information we define the matrix $D = \{d_{ij}\ i, j = 1, 2, ..., n\}$.

For the Multidimensional scaling approach we consider the node localization problem with defining the network as an undirected graph with vertices *V* and edges *E*. The vertices correspond to the nodes, of which zero or more may be special nodes, which we call anchors, whose positions are already known. We assume that all the nodes being considered in the positioning problem form a connected graph, i.e., there is a path between every pair of nodes.

We focus on classical MDS in this paper. Classical MDS is the simplest case of MDS: the proximities of objects are treated as distances in a Euclidean space. The goal of MDS is to find a configuration of points in a multidimensional space such that the inter-point distances are related to the provided proximities by some transformation (e.g., a linear transformation).

Let $p_{ij}$ refer to the proximity measure between objects $i$ and $j$. The Euclidean distance between two points $X_i = (x_{i1}; x_{i2}; x_{i3}...x_{im})$ and $X_j = (x_{j1}; x_{j2}; x_{j3}...x_{jm})$ in an $m$ - dimensional space is

$$d_{ij} = \sqrt{\sum_{k=1}^{m}(x_{ik} - x_{jk})^2} \tag{1}$$

The Euclidean distances are related to the proximities by a transformation $d_{ij} = f(p_{ij})$. In the classical MDS, a linear transformation model is assumed, i.e. $d_{ij} = a + bp_{ij}$. The distances $D$ are determined so that they are as close to the proximities $P$ as possible. In that way, we define $I(P) = D + E$, where $I(P)$ is a linear transformation of the proximities, and $E$ is a matrix of errors. Since $D$ is a function of the coordinates $X$, the goal of classical MDS is to calculate $X$ such that the sum of squares of $E$ is minimized, subject to suitable normalization of $X$.

In classical MDS, $P$ is shifted to the center and coordinates $X$ can be computed from the double centered $P$ through singular value decomposition. For an $n \times n$ $P$ matrix for $n$ points and $m$ dimensions of each point, it can be shown that

$$-\frac{1}{2}(p_{ij}^2 - \frac{1}{n}\sum_{i=1}^{n} p_{ij}^2 - \frac{1}{n}\sum_{i=1}^{n} p_{ij}^2 + \frac{1}{n^2}\sum_{i=1}^{n}\sum_{j=1}^{n} p_{ij}^2)$$
$$= \sum_{k=1}^{m} x_{ik} x_{jk}$$

(2)

The double centered matrix on the left hand side ($B$) is symmetric. Having now calculated $B$ and performing singular value decomposition on $B$ gives $B = VAV$. The coordinate matrix becomes $X = VA^{1/2}$.

Retaining the first $r$ largest eigenvalues and eigenvectors $(r < m)$ leads to a solution in lower dimension. This implies that the summation over $k$ runs from 1 to $r$ instead of $m$. This is the best low rank approximation in the least-squares sense. For example, for a 2D network, we take the first 2 largest eigenvalues and eigenvectors to construct the best 2D approximation.

**2.1 MDS-MAP algorithm**

MDS-MAP is a localization method based on multidimensional scaling [3]. The MDS-MAP algorithm consists of three steps:

- Compute the shortest distances between all pairs of nodes in the region. The computed distances are used for building the distance matrix for MDS.
- Apply classical MDS to the distance matrix, retaining the first 2 eigenvalues and eigenvectors to construct a 2D relative map.
- Given sufficient anchor nodes (3 or more), transform the relative map to an absolute map based on the absolute positions of anchors.

In the first step, we assign distances to the edges in the connectivity graph. When the distance of a pair of neighbor nodes is known, the value of the corresponding edge is the measured distance. When we only have connectivity information, a simple approximation is to assign value 1 to all edges. Then a classical all-pairs shortest-path algorithm, such as Dijkstra's algorithm, can be applied. In the second step, classical

MDS is applied directly to the distance matrix. The core of classical MDS is singular value decomposition. The result of MDS is a relative map that gives a location for each node. Although these locations may be accurate relative to one another, the entire map will be arbitrarily rotated relative to the true node positions. In the third step, the relative map is transformed through linear transformations, which include scaling, rotation, and reflection. The goal is to minimize the sum of squares of the errors between the true positions of the anchors and their transformed positions in the MDS map.

## 3 Semidefinite Programming

Semidefinite programming is the other approach we are going to present in this paper. It is a new optimization algorithm that uses techniques of linear programming.

It will be helpful to first introduce some mathematical notations to describe this technique. The trace of a given matrix $A$, denoted by $t_r(A)$ is the sum of the entries on the main diagonal of $A$. A symmetrical matrix is called semidefinite if all its eigenvalues are nonnegative and is represented by $A \succeq 0$.

Suppose two nodes $x_1$ and $x_2$ are within radio range $R$ of each other, the proximity constraint can be represented as a convex second order cone constraint of the form $\|x_1 - x_2\|_2 \leq R$, and this can be formulated as a matrix linear inequality.

$$\begin{pmatrix} I_2 R & x_1 - x_2 \\ (x_1 - x_2)^T & R \end{pmatrix} \succeq 0 \quad (3)$$

The mathematical model of the localization problem can be described as follows. There are $n-m$ distinct sensor points in $R^{\dim}$ whose locations are to be determined, and other $m$ fixed points (called the anchor points) whose locations are known. The known nodes are indicated by $a$ and the unknown nodes by $\hat{x}$, so that $X = \left[\hat{x}_1, \ldots \hat{x}_{n-m}, a_1, \ldots, a_m\right]$. All $(i,j) \in E$ where $i < j$ and if $j$ is an anchor are denoted by $N_a$, and all $(i,j) \in E$, where $i < j$ are unknown is denoted by $N_x$. The following constraints must be satisfied:

$$\begin{aligned} \|a_k - \hat{x}_j\|^2 &= d^2_{kj} \quad \forall (k,j) \in N_a \\ \|\hat{x}_i - \hat{x}_j\|^2 &= d^2_{ij} \quad \forall (i,j) \in N_x \end{aligned} \quad (4)$$

We consider the case when the node distances are known, therefore, let $X = \left[\hat{x}_1;...;\hat{x}_{n-m}\right]^T$ be the matrix in $R^{\dim \times (n-m)}$ that needs to be determined. Define $e_{ij} \in R^{n-m}$ with 1 on $i$-th position and with -1 on $j$-th position, and everywhere else zeros. If $I_{\dim}$ is the identity matrix, the constraints can be written:

$$\left(a_k;e_j\right)^T \left[I_{\dim};X^T\right]\left[I_{\dim};X\right]\left(a_k;e_j\right) = d_{kj}^2 \quad \forall (k,j) \in N_a \tag{5}$$

$$e_{ij}^T X^T X e_{ij} = d_{ij}^2 \quad \forall (i,j) \in N_x$$

We now need to find a symmetric matrix $Y \in R^{\dim \times \dim}$ and $X$ that satisfy the following constraints:

$$\left(a_k;e_j\right)^T \begin{pmatrix} I_{\dim} & X \\ X^T & Y \end{pmatrix} \left(a_k;e_j\right) = d_{kj}^2 \quad \forall (k,j) \in N_a \tag{6}$$

$$e_{ij}^T Y e_{ij} = d_{ij}^2 \quad \forall (i,j) \in N_x$$

$$Y = X^T X$$

This is the SDP formulation of the problem of wireless sensor networks localization. In [4] a relaxation of this method is proposed that we will use in our simulations. The constraint $Y = X^T X$ is relaxed with $Y \succeq X^T X$. We can write this condition as follows

$$Z = \begin{pmatrix} I_{\dim} & X \\ X^T & Y \end{pmatrix} \succeq 0 \tag{7}$$

In this way the SDP problem can be written as $\min \; 0$, such that

$$Z_{1:\dim,1:\dim} = I_{\dim} \tag{8}$$

$$(0;e_{ij})(0;e_{ij})^T \bullet Z = d_{ij}^2 \quad \forall (i,j) \in N_x$$

$$(a_k;e_j)(a_k;e_j)^T \bullet Z = d_{kj}^2 \quad \forall (k,j) \in N_a$$

$$Z \succeq 0$$

where $A \bullet B = t_r(AB)$ and $0$ the zero vector of the corresponding dimension. When we have a solution to this problem, we can then easily extract the solution for the positions of the unknown nodes, since they are then defined by $X$ and are a part of $Z$. Practically this is solved by a SDP solver such as SeDuMi which we used in our simulations.

## 4 Simulation results

In our experiments, we ran MDS-MAP and SDP algorithms on various topologies of networks in Matlab. Three different network topologies were considered: (1) random topology with a uniform distribution within a square area, (2) square grid topology with some placement errors, and (3) on a hexagonal grid topology with some placement errors.

For the SDP approach the computational results presented here were generated using the interior-point algorithm SDP solvers SeDuMi [6] with their interfaces to Matlab.

In a square grid, with a placement error, a random value drawing from a normal distribution N(0;1) is added to the node's original grid position. The placement error in a hexagonal grid topology is defined similarly.

The data points represent averages over 20 rounds in networks containing 64 nodes. The anchor nodes are selected randomly and the number of anchor nodes varies between 4 and 10 in each simulation. The connectivity (average number of neighbors) is controlled by specifying radio range R. Nodes are placed in a square area with size of rxr (r=0.5).

**4.1** Random network topology

In the case when the network has a random topology, 64 nodes are placed randomly in rxr square area (r=0.5). Figure 1 shows an example of this random placement and the results in the SDP approach are given. The radio range here is 0.15r, which leads to an average connectivity of 14.625, and the number of anchor nodes is 6. Figure 2 shows a comparison of MDS-MAP and SDP estimation errors in a random network topology with 64 nodes placed in a square area with size of rxr (r=0.5). The radio ranges (R) used are 0.15r, 0.18r, 0.2r and 0.25r, which lead to an average connectivity level of 13.31, 18.21, 21.55 and 29.75 respectively. The number of the anchor nodes used is 4, 5, 6 and 10.

Figure 2 shows a better performance of the estimated errors in the SDP approach in comparison with the MDS-MAP algorithm. The estimation errors of the SDP approach are almost 2 times smaller than the MDS estimation errors. For example, MDS gives an estimation error of 0.1302R with connectivity level of 13.3125, and in contrast SDP estimation error is 0.0721R (in a case with 4 anchors), and moreover, for connectivity level of 29 and more SDP estimation errors are less than 0.01R. Obviously when the connectivity level rises, estimation errors are getting smaller even by half for connectivity level less than 18. An interesting result is that the number of anchor nodes does not effect much on the SDP estimation errors.

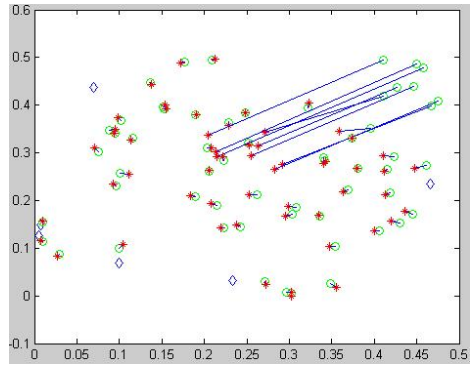

**Fig. 1.** SDP simulation of a random network topology with 64 nodes placed in a square area 0.5x0.5 and 6 anchors, average connectivity 14.625

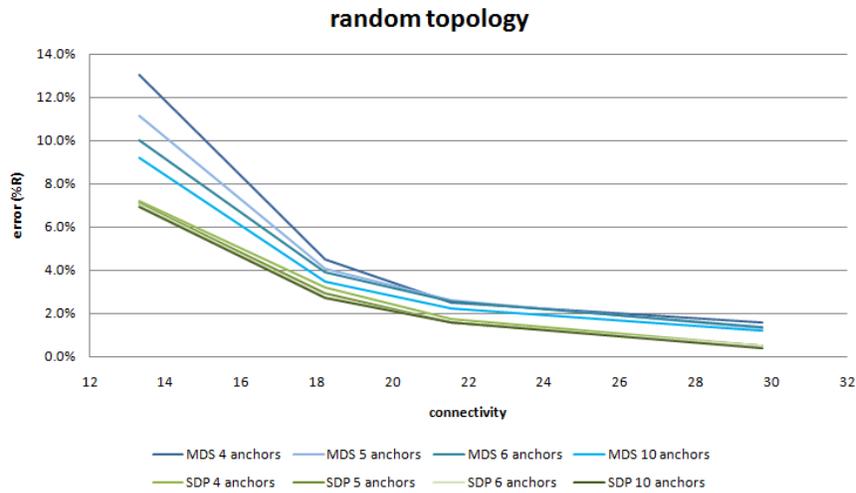

**Fig. 2.** Comparison of MDS-MAP and SDP estimation errors in a random network topology with 64 nodes placed in a square area with size of 0.5x0.5

**4.2** Square grid network topology

In the case when the network has a square grid topology, we assume that the sensor nodes are deployed according to a regular structure. Actually, nodes are placed in the neighborhood of the vertices due to random placement error. 64 nodes are placed on a rxr (r=0.5) grid, with a unit edge distance r/8. This type of network topology is shown in figure 3. The results in the SDP approach are given here. The radio range is 0.15r, which leads to an average connectivity of 13.055, and the number of anchor nodes is 6. Figure 4 shows a comparison of MDS-MAP and SDP estimation errors in a square grid network topology with 64 nodes placed in a square area with size of rxr (r=0.5). The radio ranges (R) used are 0.15r, 0.18r, 0.2r and 0.25r, which lead to a average connectivity level of 12.97, 18.07, 21.19 and 30.14 respectively.

Our results show that MDS and SDP obtain much better results on the grid layout than on the random layout for the same connectivity level. Estimation errors are lowered by half with this regular topology in comparison with the random topology. SDP outperforms MDS in the same way as in the random placement.

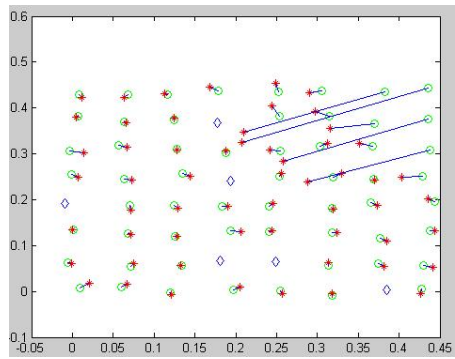

**Fig. 3.** SDP simulation of a square grid network topology with 64 nodes placed in a square area 0.5x0.5 and 6 anchors, average connectivity 13.375

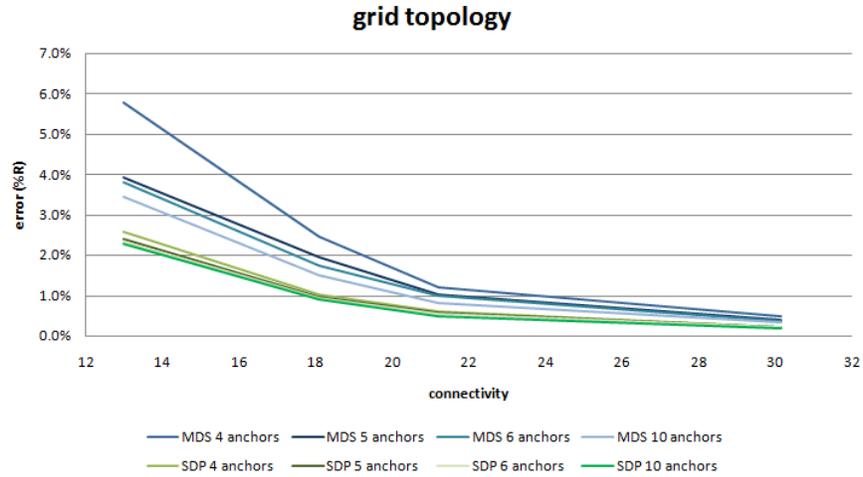

**Fig. 4.** Comparison of MDS-MAP and SDP estimation errors in a square grid network topology with 64 nodes placed in a square area with size of 0.5x0.5

**4.3** Hexagonal grid network topology

The case when the network has a hexagonal grid topology, is similar with the square grid topology. Sensor nodes are placed on the vertices of a hexagonal grid with a random placement error as in figure 5. Figure 6 shows a comparison of MDS-MAP and SDP estimation errors in a hexagonal grid network topology with 64 nodes placed in a square area with size of rxr (r=0.5). The radio ranges (R) used are 0.15r, 0.18r, 0.2r and 0.25r, which lead to a average connectivity level of 14.94, 20.35, 24.075 and 33.45 respectively. The simulation results here are similar with the square grid case. SDP estimation errors are lower than MDS estimation errors, but the improvement in the estimation errors with the SDP approach is not as stressed as in the random layout.

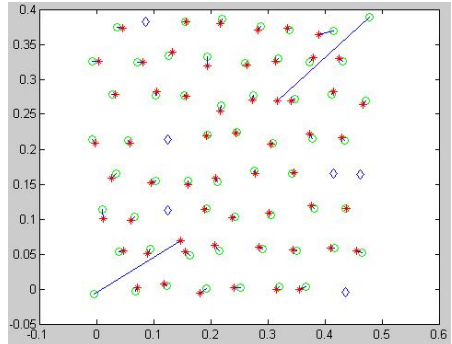

**Fig. 5.** SDP simulation of a hexagonal grid network topology with 64 nodes placed in a square area 0.5x0.5 and 6 anchors, average connectivity 14.925

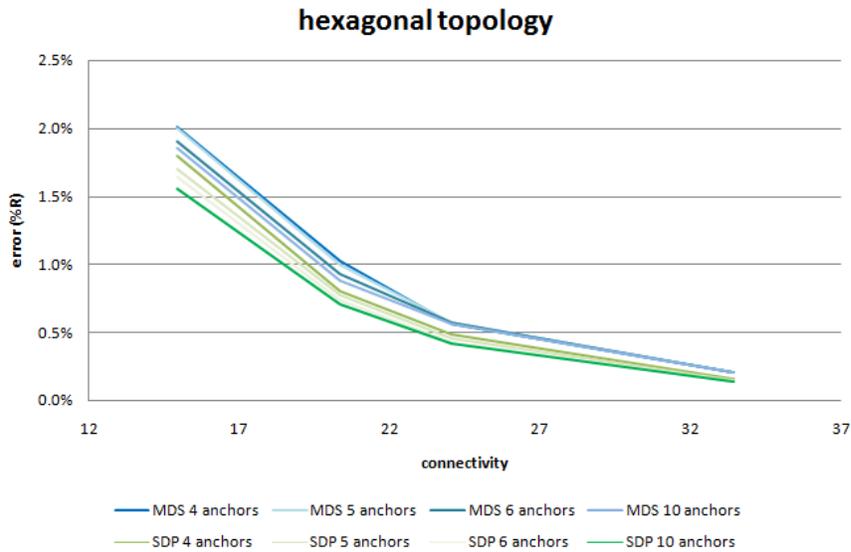

**Fig. 6.** Comparison of MDS-MAP and SDP estimation errors in a hexagonal grid network topology with 64 nodes placed in a square area with size of 0.5x0.5

## 5 Conclusions

In the vast field of research related to wireless sensor networks, our focus has been on the problem of localization, one of the major challenges in the design of ad hoc networks. Our goal was to present two approaches that are of a rising interest:

Multidimensional scaling with the MDS-MAP algorithm and localization with Semidefinite programming. They both work well, with a small amount of connectivity information about the network, however the Semidefinite programming approach with known distance network information outstands with its results in comparison to the Multidimensional Scaling approach. In conclusion, SDP as an approach is better than MDS for small sized networks (as used in our simulations) especially with random topologies. However, the time consuming factor is not considered. SDP as a more complex algorithm is slower than MDS, and for larger network sizes it will be very difficult to get any results. Although some research has been done concerning network topologies, some other irregular topologies should be considered in future with different network sizes. Furthermore, some hybrid algorithms which combine the advantages of these two approaches (greater performance with SDP and speed with MDS) should be developed.